%% file: vectorial_LBC.tex
\documentclass[review]{elsarticle}

\pdfoutput=1

\usepackage{lineno,hyperref}
\modulolinenumbers[5]

\def\picdir{pic}

\input{befehle}
\usepackage{textcomp}
\usepackage{amsmath, amssymb, amsfonts, amsthm}

\journal{Journal of Computational Physics}

\begin{document}

\begin{frontmatter}

\title{Modelling arbitrarily shaped and tightly focused laser pulses in electromagnetic codes}

\author[CELIA-Adress]{Illia Thiele}
\cortext[mycorrespondingauthor]{Corresponding author}
\ead{illia-thiele@web.de}

\author[CELIA-Adress]{Stefan Skupin}

\author[CELIA-Adress]{Rachel Nuter}

\address[CELIA-Adress]{Univ.~Bordeaux - CNRS - CEA, Centre Lasers Intenses et Applications, UMR 5107, 33405 Talence, France}

\begin{abstract}
Investigation of laser matter interaction with electromagnetic codes requires to implement sources for the electromagnetic fields. A way to do so is to prescribe the fields at the numerical box boundaries in order to achieve the desired fields inside the numerical box. Here we show that the often used paraxial approximation can lead to unexpected field profiles with strong impact on the laser matter interaction results. We propose an efficient numerical algorithm to compute the required laser boundary conditions consistent with the Maxwell's equations for arbitrarily shaped, tightly focused laser pulses.
\end{abstract}


\end{frontmatter}

\linenumbers

\section{Introduction}
Electromagnetic codes are useful tools to study various problems in microwave engineering, plasma physics, optics and other branches of natural science. Such codes solve Maxwell's equations coupled to constitutive equations describing the matter. In studies of laser matter interaction, external electromagnetic waves (the ''laser'') have to enter the computational domain in order to interact with the matter. 
In the case of particle-in-cell (PIC) codes like CALDER \cite{CALDER}, PICLS \cite{PICLS} or OCEAN \cite{PhysRevE.87.043109},
it is common practise to prescribe external electric and magnetic fields at the numerical box boundaries. 
Very often, the paraxial approximation \cite{born1999principles,goodman2005introduction} is used to calculate the required fields at the boundaries. However, the paraxial approximation is valid only if the angular spectrum of the laser pulse is sufficiently narrow. Thus, it is not possible to use this approximation for strongly focused pulses. 
For several beam types, e.g.\ Gaussian, higher order approximations have been presented \cite{Salamin2007,Sheppard:99}, but they are rather complicated and therefore not easy to implement. Moreover, for more exotic beam shapes, like vector beams or even sampled experimental profiles, it may be even impossible to find an explicit analytical solution.

In this paper, we propose a simple and efficient algorithm for a Maxwell consistent calculation of the electromagnetic fields at the boundaries of the computational domain. We call them laser boundary conditions (LBCs). Our algorithm can describe any kind of laser pulses, in particular tightly focused, arbitrarily shaped and polarized. Such laser pulses become more and more popular in the context of laser driven radiation and particle sources as well as laser material processing \cite{Yew2007453,PhysRevLett.112.215001,Buccheri:15,Cheng201388,Hnatovsky:12,kammel:lsa:3:e169,Wang2008}.

The paper is organized as follows. Section~\ref{sec:Sche_pic} details the problem we want to solve. In Sec.~\ref{sec:Las_prop}, the theory of laser propagation in vacuum is reviewed. Section~\ref{sec:LBC} describes in detail our algorithm for the computation of Maxwell consistent LBCs, and in Sec.~\ref{sec:Examp} we present two illustrative examples: A tightly focused Gaussian beam and a longitudinal needle beam. Section~\ref{sec:Conc} summarizes the results and offers perspectives for potential applications.

\section{Schematic presentation of the laser injection}
\label{sec:Sche_pic}

\begin{figure}
    \includegraphics*[width=0.5\textwidth]{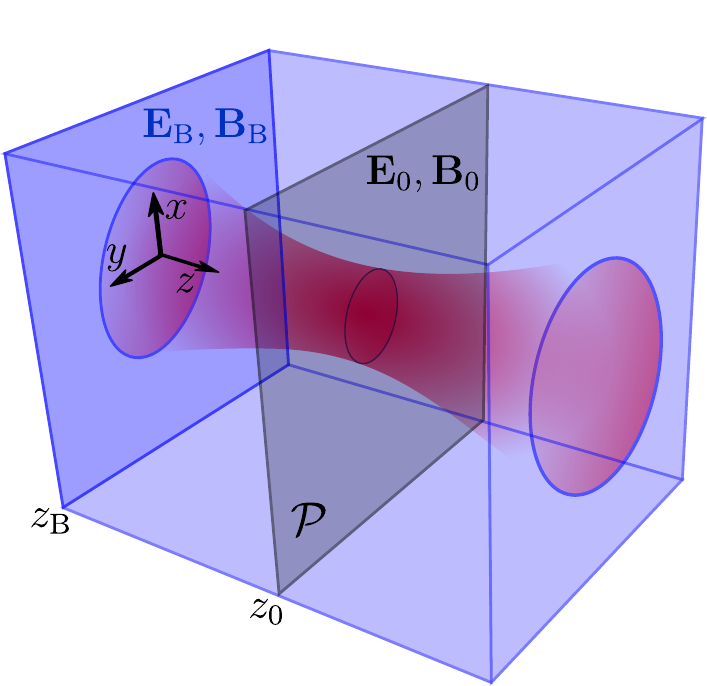}
    \centering
    \caption{Schematic picture of the laser (red) injection problem into the computational domain: Electric and magnetic fields $\Evec_0$, $\Bvec_0$ are prescribed in the plane $\mathcal{P}$ [here the $(x,y)$-plane at $z=z_0$]. The fields $\Evec_\mrm{B}$, $\Bvec_\mrm{B}$ at the boundary (blue) are unknown and have to be calculated.}
    \label{fig:intr}
\end{figure}

In numerical studies of laser matter interaction, it is common practise to define the laser by its propagation in vacuum, for example, by position and shape of the pulse at focus. 
In this paper, we choose to prescribe the pulse in a plane $\mathcal{P}$ parallel to a boundary of the rectangular numerical box, i.e., typically in the focal plane (see Fig.~\ref{fig:intr}). The laser (red) is passing through the plane $\mathcal{P}$, where the fields\footnote{Vectors are typed in bold.} $\Evec_0$, $\Bvec_0$ are prescribed for all times $t$. The goal is to calculate the fields $\Evec_\mrm{B}$, $\Bvec_\mrm{B}$ at the boundary from $\Evec_0$, $\Bvec_0$. 
As we will see in Sec.~\ref{sec:LBC}, choosing $\mathcal{P}$ parallel to a boundary allows us to resort to Fast Fourier Transforms (FFTs) in the numerical computation of the LBCs. It is of course possible to prescribe the fields in an arbitrary plane and use the general solution given in the next Section to calculate the LBCs. However, in this case one cannot exploit the advantage of an efficient computation with FFTs (Sec.~\ref{sec:LBC}) and will have to evaluate the spatial Fourier integrals directly, for example by performing discrete sums. 

\section{Laser field propagation in vacuum}
\label{sec:Las_prop}

Let $\Evec_0(\rvec_\perp,t)=\Evec(\rvec_\perp,z=z_0,t)$ and $\Bvec_0(\rvec_\perp,t)=\Bvec(\rvec_\perp,z=z_0,t)$ be the electromagnetic  fields in the plane $\mathcal{P}$. 
In the following, we want to compute $\Evec$, $\Bvec$ in the whole space and for all times. We will see that not all components of $\Evec_0$, $\Bvec_0$ can be prescribed independently. Moreover, we will comment on how to handle evanescent fields, and finally discuss the paraxial limit. 

\subsection{Propagation of electromagnetic fields and their interdependencies}
\label{sec:Prop_EB}

Electromagnetic fields in vacuum are governed by Maxwell's equations. In frequency or temporal Fourier space they read
\begin{align}
	\nabla\cdot\hat{\Evec}(\rvec, \omega)  & = 0 & \nabla\times\hat{\Evec}(\rvec, \omega) & = \rmi\omega \hat{\Bvec}(\rvec, \omega) \label{eq:Gauss_Faraday_text}\\
	\nabla\cdot \hat{\Bvec}(\rvec, \omega) & = 0 & \nabla\times\hat{\Bvec}(\rvec, \omega) & = -\rmi\omega\frac{1}{c^2} \hat{\Evec}(\rvec, \omega)\,\mbox{.} \label{eq:Gaussmag_Ampere_text}
\end{align}
Here, $\omega$ is the frequency variable, $c$ is the vacuum speed of light, and $\hat{~}$ denotes the Fourier transform with respect to time $t$.
For the definition of the Fourier transforms as used in this paper see \ref{app:FT}. The wave equation for the electric field $\Evec$ in frequency space reads (analogue for the magnetic field $\Bvec$)
\begin{equation}
	\Delta \hat{\Evec}(\rvec, \omega) + \frac{\omega^2}{c^2}\hat{\Evec}(\rvec, \omega) = 0\,\mbox{.} \label{eq:waveeq_E}
\end{equation}
Written in spatial Fourier space (with wavevector $\kvec$ as spatial Fourier variables) Eq.~(\ref{eq:waveeq_E}) would reduce to an algebraic equation and lead to the vacuum dispersion relation $\kvec^2=\omega^2/c^2$. However, we want to describe propagation of $\Evec_0$ along $z$. To this end, we keep the $z$ variable and perform the Fourier transform with respect to the transverse variables $\rvec_\perp$ only. Transforming Eq.~(\ref{eq:waveeq_E}) to transversal spatial Fourier space, where $\kvec_\perp=(k_x,k_y)^\mrm{T}$ is the transversal wavevector, gives 
\begin{equation}
	k_z^2(\kvec_\perp,\omega) \bar{\Evec}(\kvec_\perp, z, \omega) + \partial^2_z\bar{\Evec}(\kvec_\perp, z, \omega) = 0 \label{eq:waveeq_Ebar}\,\mbox{,}
\end{equation}
where $k_z(\kvec_\perp,\omega) = \sqrt{\omega^2/c^2-k_x^2 - k_y^2}$, and $\bar{~}$ denotes the temporal and transverse spatial Fourier domain. The fundamental solutions of Eq.~(\ref{eq:waveeq_Ebar}) are the forward $(+)$ and backward $(-)$ propagating, plane or evanescent waves (analogue for the magnetic field $\Bvec$)
\begin{equation}
	\bar{\Evec}^\pm	(\kvec_\perp, z, \omega) = \bar{\Evec}^\pm_0(\kvec_\perp, \omega)e^{\pm\rmi k_z(\kvec_\perp,\omega)(z-z_0)}
	\,\mbox{.}\label{eq:E_prop}
\end{equation}
It is important to note that $\Evec^\pm_0$, $\Bvec^\pm_0$ cannot be chosen arbitrarily. In fact, only two out of six vector components (for forward and backward direction, respectively) are independent. For example, we can choose to prescribe $\Evec^\pm_{0,\perp}$ in the plane $\mathcal{P}$. Then, by exploiting Eqs.~(\ref{eq:Gauss_Faraday_text}) and (\ref{eq:E_prop}), we get
\begin{align}
        \bar{\Evec}^\pm_\perp(\kvec_\perp, z, \omega) &= \bar{\Evec}^\pm_{0,\perp}(\kvec_\perp, \omega)e^{\pm\rmi k_z(\kvec_\perp,\omega)(z-z_0)}
        \label{eq:Eperp_from_Eperp}\\
	\bar{E}^\pm_z(\kvec_\perp, z, \omega) &= \mp\frac{\kvec_\perp\cdot\bar{\Evec}_{\perp}^\pm(\kvec_\perp,z, \omega)}{k_z(\kvec_\perp,\omega)}
	\label{eq:Ex_from_Eperp}\\
	\bar{\Bvec}^\pm(\kvec_\perp, z, \omega) &= \frac{1}{\omega k_z(\kvec_\perp,\omega)}\mathbb{R}^\pm(\kvec_\perp,\omega)\bar{\Evec}_{\perp}^\pm(\kvec_\perp,z, \omega)\,\mbox{,}
	\label{eq:B_from_Eperp}
\end{align}
with the matrix
\begin{equation}
	\mathbb{R}^\pm(\kvec_\perp,\omega)=\left(
	\begin{matrix}
		\mp k_x k_y & \mp\left[k_z^2(\kvec_\perp,\omega)+k_y^2\right]\\
		\pm\left[k_z^2(\kvec_\perp,\omega)+k_x^2\right] & \pm k_x k_y \\
		-k_y k_z(\kvec_\perp,\omega) &  k_x k_z(\kvec_\perp,\omega)
	\end{matrix}
	\right)\,\mbox{.}
\end{equation}
Obviously, we are imposing $k_z\neq0$, which is implicitly assumed when stating that the laser is passing through the plane $\mathcal{P}$. Thus, the laser must not have any components propagating parallel to $\mathcal{P}$. In complete analogy, one could prescribe the transverse magnetic fields $\Bvec^\pm_{0,\perp}$ in the plane $\mathcal{P}$ and exploit Eqs.~(\ref{eq:Gaussmag_Ampere_text}) to compute $\Bvec^\pm$ and $\Evec^\pm$ in the whole space.
In \ref{app:Lorentz_gauge}, we give an alternative method for computing of Maxwell consistent laser fields based on the vector potential in the Lorentz gauge. Such description can be advantageous in specific cases, for example radially polarized doughnut beams~\cite{1367-2630-8-8-133}, where only one component of the vector potential is sufficient to describe the whole laser.

\subsection{Evanescent fields and the paraxial limit}
\label{eq:ev_fields}

For $k_x^2 + k_y^2 > \omega^2/c^2$, $k_z(\kvec_\perp,\omega)$ becomes imaginary and Eq.~(\ref{eq:E_prop}) describes evanescent waves, with exponentially growing or decaying amplitude in $z$ direction. In free space propagation, evanescent waves violate energy conservation and are thus unphysical and do not exist. 
In order to get rid of evanescent waves, the spatial Fourier spectrum of $\Evec_0$ and $\Bvec_0$ has to be filtered in transverse spatial Fourier space, such that it contains only components with $k_x^2 + k_y^2 < \omega^2/c^2$. This condition is nothing else then ensuring the Abbe diffraction limit~\cite{born1999principles} for the fields prescribed at $z=z_0$, which, for instance, forbids to focus a beam to arbitrary small transverse size. 

In contrast, if the spatial Fourier spectrum of $\Evec_0$ and $\Bvec_0$ is nonzero only for $k_x^2+k_y^2\ll \omega^2/c^2$, one can expand $k_z$ as a Taylor series and approximate
\begin{equation}
k_z(\kvec_\perp, \omega) \approx \frac{|\omega|}{c} - \frac{c}{2 |\omega|}\left(k_x^2+k_y^2\right)\,\mbox{.}
\end{equation}
Then, Eqs.~(\ref{eq:Eperp_from_Eperp})--(\ref{eq:B_from_Eperp}) simplify as
\begin{align}
        \bar{\Evec}^\pm_\perp(\kvec_\perp, z, \omega) & \approx \bar{\Evec}^\pm_{0,\perp}(\kvec_\perp, \omega) e^{\pm\rmi \left[\frac{|\omega|}{c} - \frac{c}{2 |\omega|}\left(k_x^2+k_y^2\right) \right](z-z_0)}  \label{eq:lbcparax1} \\ 
	\bar{E}^\pm_z(\kvec_\perp, z, \omega) & \approx 0  \mspace{90.0mu} \bar{B}^\pm_x(\kvec_\perp, z, \omega)  \approx \mp \frac{1}{c} \bar{E}^\pm_y(\kvec_\perp, z, \omega) \\
	\bar{B}^\pm_z(\kvec_\perp, z, \omega) & \approx 0  \mspace{90.0mu} \bar{B}^\pm_y(\kvec_\perp, z, \omega)  \approx \pm \frac{1}{c} \bar{E}^\pm_x(\kvec_\perp, z, \omega)  \label{eq:lbcparax3} \,\mbox{,} 
\end{align}
which is well known as the paraxial or Fresnel approximation~\cite{goodman2005introduction}.

\section{Implementing the laser boundary conditions}
\label{sec:LBC}

Let us now describe a practical implementation of LBCs based on the solution of Maxwell's equations as derived in the previous Section. In the following, the laser will propagate in forward direction $(+)$ along $z$, i.e., we inject the laser from the left side of the box (see Fig.~\ref{fig:intr}).
We prescribe the electric field $\Evec_{0,\perp}(\rvec_\perp,t)$ in the plane $\mathcal{P}$ at $z=z_0$, for example a Gaussian profile in $t$ and $\rvec_\perp$. 
Then, we want to calculate the fields $\Evec_{\mrm{B}}(\rvec_\perp,t)$ and $\Bvec_{\mrm{B}}(\rvec_\perp,t)$ at the boundary $z=z_{\mrm{B}}$ on the numerical grid for all times. 
Let us consider an equidistant rectangular grid $x^i$, $y^j$, indices $i$, $j$ running from 1 to $N_x$, $N_y$, respectively, and with spatial resolution $\delta x$, $\delta y$. We evaluate $\Evec_{0,\perp}$ at the grid points $x^i$, $y^j$ for equidistant times $t^n$, $n$ is running from 1 to ${N_t}$, with temporal resolution $\delta t$:
\begin{equation}
	\Evec^{ijn}_{0,\perp} = \Evec_{0,\perp}(x^i, y^j, t^n)\,\mbox{.}
	\label{eq:discretization}
\end{equation}
The following algorithm computes the electric and magnetic fields ${\Evec}^{ij}_{\mrm{B}}(t)$ and ${\Bvec}^{ij}_{\mrm{B}}(t)$ at the boundary $z=z_\mrm{B}$ for any given time $t\in[t^1-\frac{z_\mrm{B}-z_0}{c},t^{N_t}-\frac{z_\mrm{B}-z_0}{c}]$:
\begin{enumerate}
	\item Calculate $\hat{\Evec}^{ijn}_{0,\perp}$ via discrete Fourier transforms (DFTs) in time~\cite{Press:1992:NRC:148286}:
	\begin{align}
	\omega^n & = \frac{2 \pi}{N_t \delta t}\left( -\frac{N_t}{2} + n \right) \\
	\hat{\Evec}^{ijn}_{0,\perp} & = \frac{\delta t}{2 \pi} \sum \limits_{l=1}^{N_t} \Evec^{ijl}_{0,\perp} e^{\rmi{\omega^n} t^l}\,\mbox{,} \qquad n=1, \ldots, N_t \,\mbox{.} 
	\end{align}
	\item Calculate $\bar{\Evec}^{ijn}_{0,\perp}$ via two-dimensional DFTs in transverse space:
	\begin{align}
	k_x^i & = \frac{2 \pi}{N_x \delta x}\left( -\frac{N_x}{2} + i \right) \mspace{54.0mu} k_y^j = \frac{2 \pi}{N_y \delta y}\left( -\frac{N_y}{2} + j \right) \\
	\bar{\Evec}^{ijn}_{0,\perp} & = \frac{\delta x \delta y}{(2 \pi)^2} \sum\limits_{l,m=1}^{N_{x},N_{y}} \hat{\Evec}^{lmn}_{0,\perp}e^{-\rmi(k_x^i x^l + k_y^j y^m)}
	\,\mbox{,} \qquad i,j=1, \ldots, N_{x,y} \,\mbox{.} 
	\end{align}	
	\item Calculate transverse electric field components at the boundary ($z=z_{\mrm{B}}$):
	\begin{align}
	k_z^{ijn} & = \Re \sqrt{\frac{(\omega^n)^2}{c^2} - (k_x^i)^2 - (k_y^j)^2} \label{eq:ks_discrete}\\
	\bar{\Evec}_{\mrm{B},\perp}^{ijn} & = 		\begin{cases}
			\bar{\Evec}_{0,\perp}^{ijn} e^{\rmi k_z^{ijn}(z_\mrm{B}-z_0)} & \mrm{for}~k_z^{ijn}>0 \\
			0 & \mrm{for}~k_z^{ijn}=0
		\end{cases} \,\mbox{.} \label{eq:ep_prop}
	\end{align}
	Here, $\Re$ denotes the real part of a complex number. Note that we have set $k_z^{ijn}\equiv 0 $ and $\bar{\Evec}_{\mrm{B},\perp}^{ijn}\equiv 0$ for indices $i,j,n$ with $(k_x^i)^2 + (k_y^j)^2 \geq (\omega^n)^2/c^2$, in order to suppress evanescent waves (see Sec.~\ref{eq:ev_fields}). 
	\item Calculate the longitudinal electric field component at $z=z_{\mrm{B}}$:
	\begin{equation}
	E^{ijn}_{\mrm{B},z} =
		\begin{cases}
			-\frac{{k_x^i}E^{ijn}_{\mrm{B},x} + {k_y^j}E^{ijn}_{\mrm{B},y}}{k_z^{ijn}} & \mrm{for}~k_z^{ijn}>0 \\
			0 & \mrm{for}~k_z^{ijn}=0
		\end{cases} \,\mbox{.} 
	\end{equation}
	\item Calculate the magnetic field at $z=z_{\mrm{B}}$:
	\begin{align}
		\mathbb{R}^{ijn} & =\left(
		\begin{matrix}
			-{k_x^i} {k_y^j} & (k_x^i)^2-\frac{(\omega^n)^2}{c^2}\\
			\frac{(\omega^n)^2}{c^2}-(k_y^j)^2 & {k_x^i} {k_y^j} \\
			-{k_y^j} k_z^{ijn} &  {k_x^i} k_z^{ijn}
		\end{matrix}
		\right) \\
		\bar{\Bvec}_{\mrm{B}}^{ijn} & = 
		\begin{cases}
			\frac{1}{\omega^n k_z^{ijn}}\mathbb{R}^{ijn}\bar{\Evec}^{ijn}_{\mrm{B},\perp} & \mrm{for}~k_z^{ijn}>0 \\
			0 & \mrm{for}~k_z^{ijn}=0
		\end{cases} \,\mbox{.} 
	\end{align}
	\item Calculate $\hat{\Evec}^{ijn}_{\mrm{B}}$ and $\hat{\Bvec}^{ijn}_{\mrm{B}}$ via two-dimensional inverse DFTs:
	\begin{align}
	\hat{\Evec}^{ijn}_{\mrm{B},\perp} & = \frac{(2 \pi)^2}{N_x N_y \delta x \delta y} \sum\limits_{l,m=1}^{N_{x},N_{y}} \bar{\Evec}^{lmn}_{\mrm{B}}e^{\rmi(k_x^l x^i + k_y^m y^j)} \\
	\hat{\Bvec}^{ijn}_{\mrm{B},\perp} & = \frac{(2 \pi)^2}{N_x N_y \delta x \delta y} \sum\limits_{l,m=1}^{N_{x},N_{y}} \bar{\Bvec}^{lmn}_{\mrm{B}}e^{\rmi(k_x^l x^i + k_y^m y^j)} \,\mbox{.} 
	\end{align}	
	\item Calculate ${\Evec}^{ij}_{\mrm{B}}(t)$ and ${\Bvec}^{ij}_{\mrm{B}}(t)$ for any given time $t\in[t^1-\frac{z_\mrm{B}-z_0}{c},t^{N_t}-\frac{z_\mrm{B}-z_0}{c}]$:
	\begin{align}
	{\Evec}^{ij}_{\mrm{B},\perp}(t) & = \frac{(2 \pi)}{N_t \delta t} \sum\limits_{n=1}^{N_{t}} \hat{\Evec}^{ijn}_{\mrm{B}}e^{-\rmi{\omega^n} t} \label{eq:eb}\\
	{\Bvec}^{ij}_{\mrm{B},\perp}(t) & = \frac{(2 \pi)}{N_t \delta t} \sum\limits_{n=1}^{N_{t}} \hat{\Bvec}^{ijn}_{\mrm{B}}e^{-\rmi{\omega^n} t} \label{eq:bb}\,\mbox{.} 
	\end{align}	
\end{enumerate}
The DFTs in steps 1, 2, and 6 can be calculated efficiently by means of FFTs. There are various FFT libraries available, one of the most popular and efficient implementations is the FFTW~\cite{fftw}. One has to take into account the particular definitions of spatial and temporal Fourier transform used in this paper (see \ref{app:FT}), as well as the conventions of the particular FFT library. For the FFTW~\cite{fftw}, one has to use the forward transform (flag $\tt{FFTW\_FORWARD}$) in step 2, and the backward transform (flag $\tt{FFTW\_BACKWARD}$) in steps 1 and 6. 

\begin{figure}[t]
    \includegraphics*[width=0.8\textwidth]{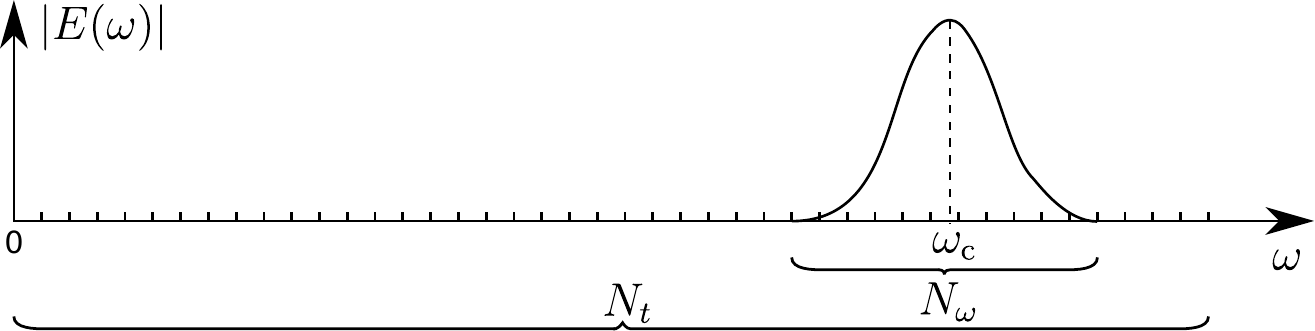}
    \centering
    \caption{Sketch of the electric field amplitude for a multi-cycle laser pulse in frequency domain. The spectrum is significantly different from zero only in $N_{\omega} \ll N_t$ frequency points.}
    \label{fig:sampling}
\end{figure}

The Fourier sums in step 7 allow to compute ${\Evec}^{ij}_{\mrm{B}}(t)$ and ${\Bvec}^{ij}_{\mrm{B}}(t)$ for any given time $t$ by means of discrete Fourier interpolation.
In fact, most of the discrete frequencies $\omega^n$ will have a negligible contribution to the spectrum when we are dealing with not-too-short laser pulses, i.e., a pulse envelope modulated with the centre frequency $\omega_\mrm{c}$ (see Fig.~\ref{fig:sampling}). By taking only the significant summands into account when evaluating the Fourier sums Eqs.~(\ref{eq:eb}) and (\ref{eq:bb}) reduces significantly both memory consumption and execution time.

When using DFTs to approximate continuous Fourier transforms as in the proposed algorithm above, one has to be careful with respect to sampling rates and the inevitable periodic boundary conditions. The initial datum $\Evec_{0,\perp}$ has to be well resolved in space and time, and one has to check that the beam fits well in the transverse numerical box for all relevant $z$ (e.g., the beam width may be larger at the boundary $z=z_{\mrm{B}}$ due to diffraction). Finally, one should not forget that Eqs.~(\ref{eq:eb}) and (\ref{eq:bb}) should be evaluated for times $t$ in the interval $[t^1-\frac{z_\mrm{B}-z_0}{c},t^{N_t}-\frac{z_\mrm{B}-z_0}{c}]$ only, otherwise a pulse train will be injected due to periodicity in time.

In a practical implementation, steps 1-6 will be performed by a pre-processor before launching the main simulation. Then, only the relevant (nonzero) contents of the arrays $\hat{\Evec}^{ijn}_{\mrm{B}}$ and $\hat{\Bvec}^{ijn}_{\mrm{B}}$ (see remark above) will be passed to the main code and step 7 will be calculated at each time step of the main simulation.

Before going on with examples, we want to make a last remark concerning the grid structure of particular Maxwell solvers. For solvers like the "Directional Splitting scheme"~\cite{DSNuter}, $\Evec$ and $\Bvec$ are discretized on the same equidistant grid and the above algorithm can be applied directly. For other solvers, like the "Yee scheme" \cite{Yee}, the fields are described on grids shifted by $\delta_x/2$, $\delta_y/2$, $\delta_z/2$, respectively. In such case, a straight forward work around would be to run the pre-processor several times with transversely shifted grids and/or shifted boundary, in order to compute the desired field components for laser injection.

\section{Examples}
\label{sec:Examp}

\subsection{Tightly focused Gaussian pulse}
\label{sec:Gauss}
Tightly focused pulses are potentially interesting for various kinds of experiments giving the possibility to achieve high intensities at rather low pulse energy or to generate micro-plasmas. Here, we are going to simulate a tightly focused Gaussian pulse and its interaction with an initially neutral gas, that is going to be ionized during the interaction. The electromagnetic fields resulting from LBCs in paraxial approximation Eqs.~(\ref{eq:lbcparax1})-(\ref{eq:lbcparax3}), as they are often applied in PIC codes, will be compared with LBCs according to the Maxwell consistent approach Eqs.~(\ref{eq:Eperp_from_Eperp})-(\ref{eq:B_from_Eperp}). For sake of computational costs, we restrict ourselves to the two dimensional case, where $\partial_y \equiv 0$ accounts for translational invariance in transverse $y$ direction.
For both cases a linear polarized Gaussian pulse is prescribed in the focal plane $z=z_0$ by 
\begin{equation}
	\Evec_{0,\perp}(x,t) = E_0 e^{-\left(\frac{x}{w_0}\right)^2-\left(\frac{t}{t_0}\right)^2}\cos(\omega_\mrm{c}t)\evec_x\,\mbox{,} 
	\label{eq:A_gaus}
\end{equation}
with center wavelength $2 \pi c/\omega_c = \lambda_\mrm{c} = 0.8$~\textmu m, pulse duration $t_0 = 20$~fs, peak intensity $I_0 = \epsilon_0 c \vert E_0\vert^2/2=5 \times 10^{14}~\mbox{W/cm}^2$ giving $E_0 = 61.4~\mbox{GV/m}$ and beam width $w_0 = 0.35$~\textmu m. The particular choice of the beam width $w_0$ implies that non-negligible parts of $\bar{\Evec}_{0,\perp}(k_x,\omega)$ are evanescent. 
These evanescent fields are suppressed in the calculation of $\Evec_{\mrm{B}}(\rvec_\perp,t)$ and $\Bvec_{\mrm{B}}(\rvec_\perp,t)$ at the boundary $z=z_{\mrm{B}}$ fully compatible with Abbe's diffraction limit (see Sec.~\ref{eq:ev_fields}). This leads to a 10\% larger full-width-at-half-maximum (FWHM) beam width and smaller electric field at focus.

\begin{figure}[t]
    \includegraphics*[width=1.0\textwidth]{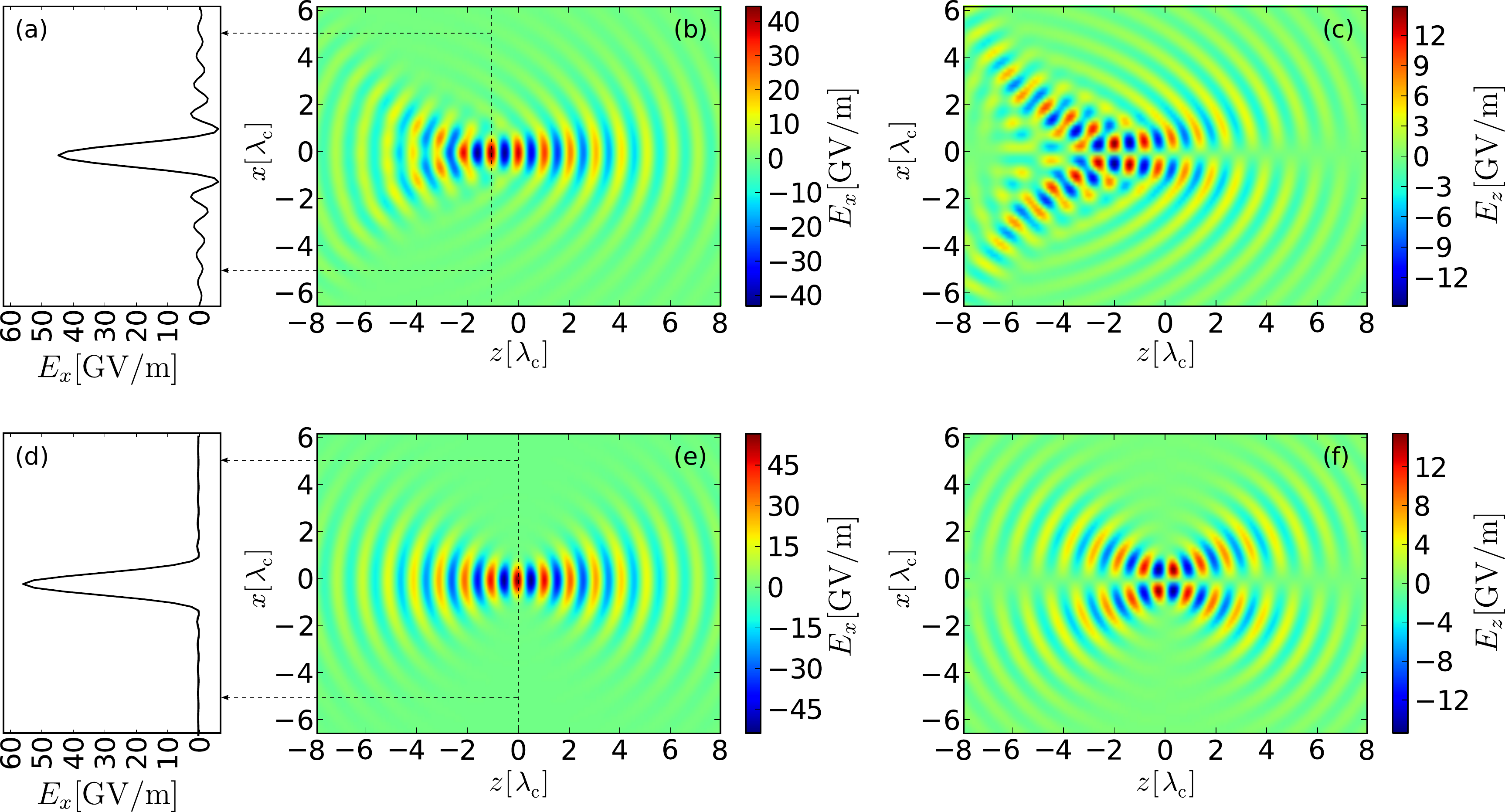}
    \centering
    \caption{Comparison of LBCs in paraxial approximation Eqs.~(\ref{eq:lbcparax1})-(\ref{eq:lbcparax3}) (a-c) and according to the Maxwell consistent approach Eqs.~(\ref{eq:Eperp_from_Eperp})-(\ref{eq:B_from_Eperp}) (d-e). Snapshots of transversal fields $E_x$ (b,e) and longitudinal fields $E_z$ (c,f) of a tightly focused Gaussian pulse (see text for details) reveal strong distortions in case of the paraxial LBCs. Calculations were performed using the PIC code OCEAN \cite{PhysRevE.87.043109}, assuming an argon atmosphere at ambient pressure. In (a) and (d) line-outs of the transversal electric field $E_x$ at focus are presented, revealing strong side-wings in the beam profile for the paraxial LBCs. The laser propagates from left to right.}
    \label{fig:fields_gaus}
\end{figure}

We solve Maxwell's equations numerically using the PIC code OCEAN~\cite{PhysRevE.87.043109}. In all simulations we consider an argon atmosphere at ambient pressure. 
Figure~\ref{fig:fields_gaus} compares snapshots of transversal ($E_x$) and longitudinal ($E_z$) electric field components for paraxial (a-c) and Maxwell consistent (d-f) LBCs when the pulse is at focus. Distortions in the fields produced by the paraxial LBCs [see Fig.~\ref{fig:fields_gaus}(b,c)] are clearly visible, even the focus (position of smallest beam width) is shifted by more than 1~\textmu m from the expected position at $z_0=0$~\textmu m. Both transversal and longitudinal field amplitudes are not symmetric with respect to the focus. As the line-out at focus in Fig.~\ref{fig:fields_gaus}(a) shows, non-negligible side-wings appear outside the main lobe. In contrast, the Maxwell consistent LBCs produce symmetric fields [see Fig.~\ref{fig:fields_gaus}(e,f)] with respect to the focus at $z_0=0$, and the line-out in Fig.~\ref{fig:fields_gaus}(d) shows no side-wings in the beam profile. The maximum transversal electric field amplitude for the paraxial LBCs is significantly lower than that achieved with the Maxwell consistent LBCs.
For both LBCs, the longitudinal field amplitude reaches about 30\% of the transversal field amplitude, a direct consequence of the steep transversal gradients in the beam profile. 

\begin{figure}[t]
    \includegraphics*[width=0.8\textwidth]{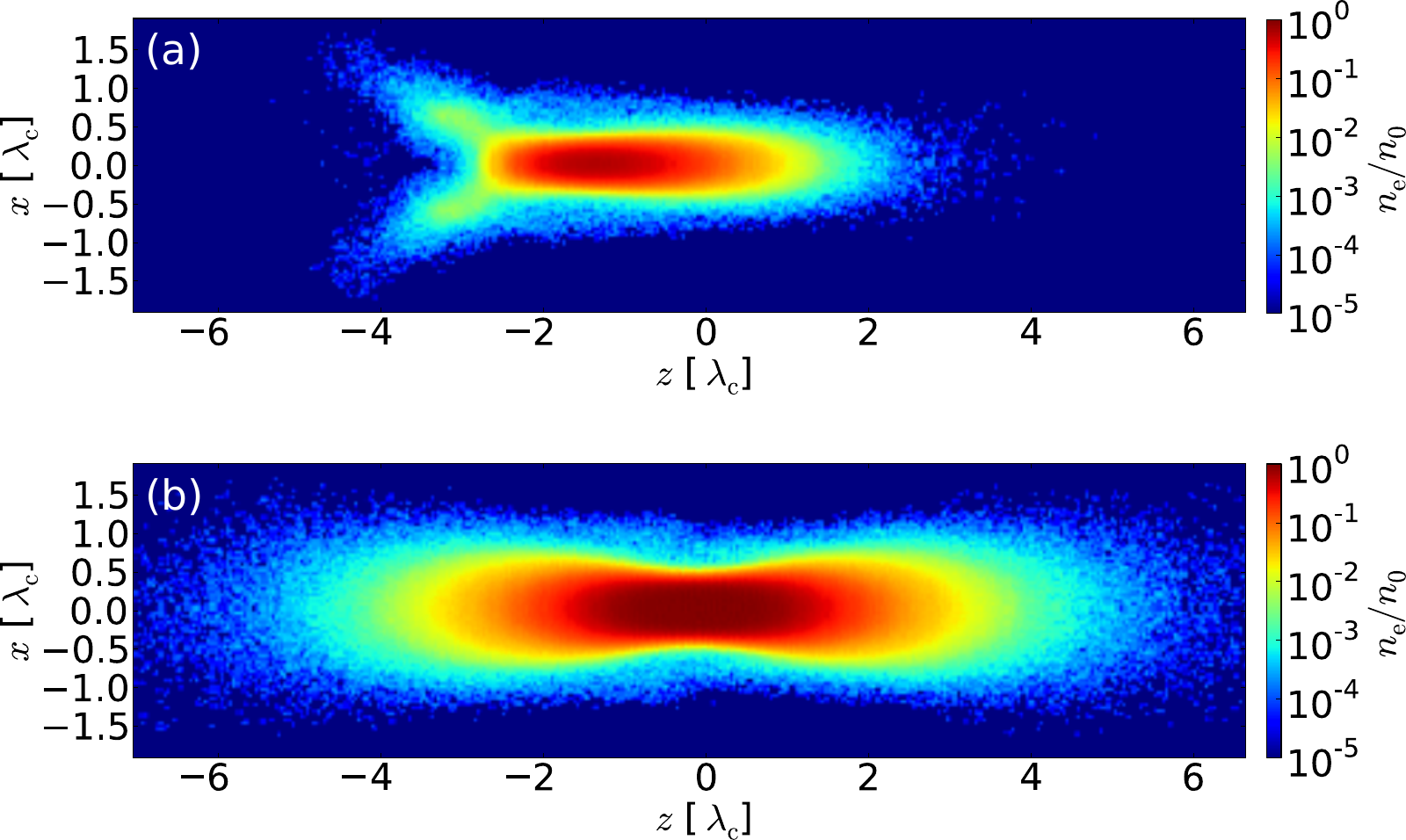}
    \centering
    \caption{Electron densities $n_\mrm{e}$ produced by the tightly focused Gaussian laser pulses shown in Fig.~\ref{fig:fields_gaus} (see text for details). The profile produced by paraxial LBCs (a) is even qualitatively different than the one produced by Maxwell consistent LBCs (b). Electron densities are scaled to the initial neutral density $n_0$. The laser propagates from left to right.}
    \label{fig:density_gaus}
\end{figure}

The code OCEAN fully accounts for ionization according to the quasistatic ADK theory \cite{Nuter11,Ammosov-1986-Tunnel,PhysRevA.64.013409} and uses  ionization data from~\cite{Carlson197063}. It is thus instructive to inspect the electron plasma generated by the tightly focused laser pulses for paraxial and Maxwell consistent LBCs. The resulting distributions of the electron density $n_\mrm{e}$ after the laser pulse has passed through the interaction region are shown in Fig.~\ref{fig:density_gaus}.
The electron density profiles are even qualitatively different for paraxial and Maxwell consistent LBCs: The paraxial LBCs give a fish-like shape, where before the focus (negative $z$) the peak electron density appears off-axis [see Fig.~\ref{fig:density_gaus}(a)], and only up to 60\% of the argon atoms get ionized. In contrast, the Maxwell consistent LBCs produce a cigar like shape with the peak electron density on the optical axis [see Fig.~\ref{fig:density_gaus}(b)], and a fully ionized plasma is produced. We would like to stress that these deviations in the plasma profile are far from negligible, and may have significant impact on features like back-reflected radiation or energy deposition in the medium. The observed sensitivity towards the LBC for tight focusing is not limited to ultrashort low energy pulses interacting with gaseous media, but should be equally important for solid targets and higher pulse energies.

\subsection{Longitudinal needle beam}
\label{sec:Vortex}

\begin{figure}[t]
    \includegraphics*[width=1.0\textwidth]{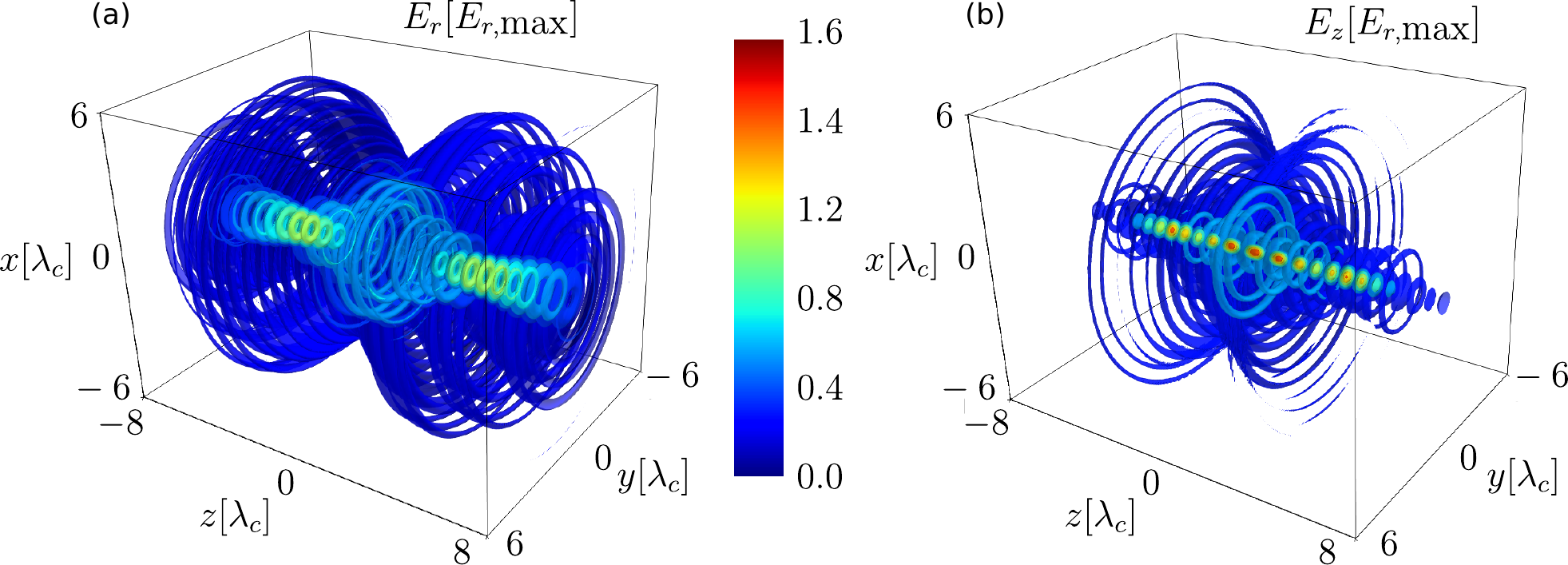}
    \centering
    \caption{The absolute values of the radial and longitudinal electric fields $E_r$ (a) and $E_z$ (b) of a longitudinal needle beam in the focal region. The fields are normalized to the maximum $E_{r,\mbox{max}}$ of the radial field $E_r$ in the whole space.}
    \label{fig:needle}
\end{figure}

In order to demonstrate generality and ease of use of the proposed Maxwell consistent LBCs, let us have a look at a (on the first glance) more complicated example. In~\cite{Wang2008}, the authors describe the ''creation of a needle of longitudinally polarized light'' by tight focusing of a radially polarized Bessel-Gaussian beam. The radial component of the electric field of such beam at focus reads
\begin{equation}
\begin{split}
	\Evec_{0,\perp}(r,t) & = \int\limits_0^{\alpha} \! T(\theta) \sqrt{\cos\theta}\sin(2\theta) e^{-\left(\frac{\sin\theta}{\sin\alpha}\right)^2}
	J_1\!\left(2\frac{\sin\theta}{\sin\alpha}\right) J_1\!\left(\frac{\omega_\mrm{c}}{c}r\sin\theta\right)d\theta \\
	& \quad \times E_0 \cos(\omega_\mrm{c}t) \evec_r \,\mbox{.}
\end{split}
	\label{eq:A_Vortex}
\end{equation}
Here, the electric field is written in cylindrical coordinates $(r,\phi,z)$, and $\evec_r$ is the radial unit vector. 
The beam profile is given as an integral over the angle $\theta$, where $\alpha$ denotes the acceptance angle of the focusing optic. Following~\cite{Wang2008}, we assume a numerical aperture $\mrm{NA}=0.95$, corresponding to $\alpha\approx 0.4 \pi$. $J_1(x)$ denotes the corresponding Bessel function. The transmission function $T(\theta)$ takes into account a binary-phase optical element, which may further increase the relative longitudinal field strength as well as the length of the needle, however, to the detriment of the optical efficiency. Here, we consider a five-belt optical element~\cite{Wang2008}
\begin{equation}
T(\theta)= 	\begin{cases}
			1 & \mrm{for}~0\le\theta < \theta_1,~\theta_2 \le \theta < \theta_3, ~\theta_4 \le \theta < \alpha\\
			-1 & \mrm{for}~\theta_1 \le\theta < \theta_2,~\theta_3 \le \theta < \theta_4
		\end{cases} \,\mbox{,} 
\end{equation}
with $\theta_1=0.0275 \pi$, $\theta_2=0.121 \pi$, $\theta_3=0.19 \pi$, and $\theta_4=0.26 \pi$.
As in the previous example, we consider a laser wavelength of $\lambda_\mrm{c}=0.8$~\textmu m. 

Figure~\ref{fig:needle} presents radial and longitudinal electric fields of the longitudinal needle beam from simulations using OCEAN~\cite{PhysRevE.87.043109} and Maxwell consistent LBCs. In agreement with~\cite{Wang2008} we find a longitudinal field amplitude that exceeds the radial one in the focal region along several laser wavelengths ($\sim 8\lambda_\mrm{c}$). The maximum longitudinal field amplitude is about 1.6 times larger than the radial one, which achieves its maximum out of focus at $z=\pm 4\lambda_\mrm{c}$. This allows the longitudinal field to dominate in the focal plane by a factor of 2.5.

\section{Conclusion}
\label{sec:Conc}

Injecting laser pulses into Maxwell solvers requires to prescribe the electromagnetic fields at the boundaries of the numerical box. Often, these fields are calculated by using the paraxial approximation. We have shown that for tightly focused beams this approach does not give the expected results. Instead, Maxwell's equations in vacuum have to be solved rigorously in order to find the proper fields at the boundaries. 
We proposed an easy to implement algorithm to achieve this goal, which allows to calculate the LBCs from transversal electric or magnetic field components defined in a plane, e.g., the focal plane. The presented algorithm can be parallelized in a straight forward manner and may be used with simulations tools employing domain decomposition.

We successfully employed our approach to simulate a tightly focused Gaussian pulse. An accurate handling of the laser injection turns out to be crucial: Electron density profiles from ionization of neutral argon atoms due to field ionization are shown to be strongly dependent on the LBCs. Consequently, the LBCs may have significant impact on features like back-reflected radiation or energy deposition in the medium. 
Furthermore, our algorithm offers a simple way to simulate more complex pulse configurations or even sampled experimental beam profiles. 
Such ''structured light'' receives a lot of recent interest from various communities~\cite{osa_structured_light}. 
As an example we demonstrated a longitudinal needle beam, which may be interesting for, among others, laser based material processing or particle acceleration studies. Thus, we believe that our approach will be useful for a larger community working on electromagnetic simulation codes.

\section*{Acknowledgements}
Numerical simulations were performed using computing
resources at M\'eso\-centre de Calcul Intensif Aquitain (MCIA), Grand
Equipement National pour le Calcul Intensif (GENCI, grant
no.~2015-056129), and Partnership for Advanced Computing in Europe
(PRACE, grant no.~2014112576).

\appendix
\section{The Fourier transforms}
\label{app:FT}

We define the temporal Fourier transform $\hat{f}(\rvec, \omega)$ of a function $f(\rvec, t)$ by
\begin{align}
	\hat{f}(\rvec, \omega) &= \frac{1}{2\pi}\int f(\rvec, t)e^{\rmi\omega t}\,dt\label{eq:IFT}\\
	f(\rvec, t) &= \int \hat{f}(\rvec, \omega)e^{-\rmi\omega t}\,d\omega \label{eq:FT}\,\mbox{.}
\end{align}
Further one, we define the transverse spatial Fourier transform $\bar{f}(\rvec_\perp, z, \omega)$ of a function $\hat{f}(\rvec, \omega)$ by
\begin{align}
	\bar{f}(\kvec_\perp, z, \omega) &= \frac{1}{(2\pi)^2}\iint \hat{f}(\rvec_\perp,z, \omega)e^{-\rmi\kvec_\perp\cdot\rvec_\perp}\,d^2\rvec_\perp\label{eq:FT2}\\
	\hat{f}(\rvec_\perp, z, \omega) &= \iint \bar{f}(\kvec_\perp,z, \omega)e^{\rmi\kvec_\perp\cdot\rvec_\perp}\,d^2\kvec_\perp\label{eq:space_FT}\,\mbox{,}
\end{align}
where $\rvec_\perp = (x,y)^\mrm{T}$ and $\kvec_\perp = (k_x,k_y)^\mrm{T}$. 

Note the difference in the sign of the exponent for temporal and spatial transform, which is common practise in the optical context.
In particular when one wants to approximate Fourier integrals by finite sums, and resort to discrete Fourier transformations (DFTs) or even fast Fourier transforms (FFTs)~\cite{Press:1992:NRC:148286},
it is important to keep track of these sign conventions (see Sec.~\ref{sec:LBC}).

\section{Generating Maxwell consistent solutions using the vector potential in Lorentz gauge}
\label{app:Lorentz_gauge}

Introducing electromagnetic potentials $\Avec$, $\phi$ in Lorrentz gauge via
\begin{gather}
	\hat{\Bvec} = \nabla \times \hat{\Avec} \mspace{90.0mu}
	\hat{\Evec} = \rmi\omega\hat{\Avec}-\nabla\hat{\phi} \label{eq:Gauge_cond} \\
	\nabla\cdot\hat{\Avec}(\rvec, \omega) = \rmi\omega\frac{1}{c^2}\hat{\phi}(\rvec, \omega)
	\label{eq:Lorentz_condition}\,\mbox{,}
\end{gather}
leads to decoupling of $\phi$ and the components of $\Avec$, and one finds (in vacuum)~\cite{jackson1999classical}
\begin{equation}
	k_z(\kvec_\perp,\omega) \bar{\Avec}(\kvec_\perp, z, \omega) + \partial^2_z\bar{\Avec}(\kvec_\perp, z, \omega) = 0\,\mbox{.} \label{eq:A_in_k_perp}
\end{equation}
In analogy to Eq.~(\ref{eq:waveeq_Ebar}), fundamental solutions are the forward $(+)$ and backward $(-)$ propagating, plane or evanescent waves
\begin{equation}
	\bar{\Avec}^\pm	(\kvec_\perp, z, \omega) = \bar{\Avec}^\pm_0(\kvec_\perp, \omega)e^{\pm\rmi k_z(\kvec_\perp,\omega)(z-z_0)}\,\mbox{.}
	\label{eq:A_prop}
\end{equation}
By plugging Eq.~(\ref{eq:A_prop}) into Eq.~(\ref{eq:Gauge_cond}), and using Eq.~(\ref{eq:Lorentz_condition}) to eliminate $\phi$, electric and magnetic fields can be expressed in terms of the vector potential at $z=z_0$:
\begin{align}
	\bar{\Bvec}^\pm(\kvec_\perp, z, \omega) & = \rmi \kvec^\pm(\kvec_\perp,\omega) \times \bar{\Avec}^\pm_0(\kvec_\perp, \omega)e^{\pm\rmi k_z(\kvec_\perp,\omega)(z-z_0)}
	\label{eq:B_of_A0}\\
	\bar{\Evec}^\pm(\kvec_\perp, z, \omega) & = \rmi\omega \left(1 - \frac{c^2}{\omega^2}\kvec^\pm(\omega)\kvec^\pm(\omega)^\mrm{T}\right)\bar{\Avec}^\pm_0(\kvec_\perp, \omega)e^{\pm\rmi k_z(\kvec_\perp,\omega)(z-z_0)}\,\mbox{.}
	\label{eq:E_of_A0}
\end{align}
In general, the three components of $\Avec^\pm_0$ can be chosen independently, however, only two components are necessary to prescribe an arbitrary laser pulse~\footnote{As shown in Sec.~\ref{sec:Prop_EB}, only two electric or magnetic field components can be set independently for a laser ($k_z\neq 0$), the corresponding divergence equation determines the third one. Hence, only two components of $\Avec^\pm_0$ are sufficient to prescribe an arbitrary laser pulse.}. The use of the vector potential can be nevertheless advantageous, because certain beams, like radially polarized doughnut beams~\cite{1367-2630-8-8-133}, can be described by a single (longitudinal) component of the vector potential. 


\bibliography{mybibfile}

\end{document}

%% file: befehle.tex
\newcommand{\mrm}[1]{\mathrm{#1}}

\newcommand{\rmi}{\mathrm{i}} 

\newcommand{\Avec}{\mathbf{A}}

\newcommand{\Evec}{\mathbf{E}}	
\newcommand{\Bvec}{\mathbf{B}}	

\newcommand{\rvec}{\mathbf{r}}
\newcommand{\evec}{\mathbf{e}}	

\newcommand{\kvec}{\mathbf{k}}

\DeclareSymbolFont{lettersA}{U}{txmia}{m}{it}
\DeclareMathSymbol{\real}{\mathord}{lettersA}{"92} 
\DeclareMathSymbol{\cplx}{\mathord}{lettersA}{"83} 

